\input harvmac

\def\half{{1\over 2}}

\def\sh{\hat{\sigma}}

\def\la{{\Lambda}}

% \def\lt{\tilde{\lambda}}

%%%%%%%%%%%%%%%%%%%%%%%%%%%%%%%%%%%%%%%%%%%%%%%%%%%%%%%%%%%%%%%%%%%%%

\Title{}{\vbox{\centerline{A Model of Holographic Dark Energy}}}

\centerline{ Miao Li } \vskip .5cm \centerline{\it Institute of
Theoretical Physics} \centerline{\it Academia Sinica, P.O. Box
2735} \centerline{\it Beijing 100080, China} \centerline{\it and}
\centerline{\it Interdisciplinary Center of Theoretical Studies}
\centerline{\it Academia Sinica, Beijing 100080, China}
\centerline{\tt mli@itp.ac.cn}

\bigskip

A model for holographic dark energy is proposed, following the
idea that the short distance cut-off is related to the infrared
cut-off. We assume that the infrared cut-off relevant to the dark
energy is the size of the event horizon. With the input
$\Omega_\Lambda=0.73$, we predict the equation of state of the
dark energy at the present time be characterized by $w=-0.90$. The
cosmic coincidence problem can be resolved by inflation in our
scenario, provided we assume the minimal number of e-foldings.

\Date{March, 2004}

%J.~D.~Bekenstein, Phys.\ Rev.\ D {\bf 7} (1973) 2333; Phys.\ Rev.\
%D {\bf 9} (1974) 3292; {\bf 23} (1981) 287.

%\bibitem{ccp} S.~Weinberg, Rev.Mod.Phys.{\bf 61}:1-23,1989.

\nref\snc{A. G. Riess et al., Astron. J. 116 (1998) 1009; S.
Perlmutter et al., APJ 517 (1999) 565.}
\nref\ben{C. L. Bennett et
al., astro-ph/0302207; D. N. Spergel et al., astro-ph/0302209; H.
V. P. Peiris et al., astro-ph/0302225.}
\nref\ckn{A. Cohen,
D.~Kaplan and A.~Nelson, hep-th/9803132, Phys. Rev.  Lett. 82
(1999) 4971.}
\nref\st{P. Horava and D. Minic,
hep-th/hep-th/0001145, Phys.Rev.Lett. 85 (2000) 1610; S. Thomas,
Phys. Rev. Lett.  89 (2002) 081301.}
\nref\sh{S. D. H. Hsu,
hep-th/0403052.}
\nref\nsnc{ A. G. Riess et al, astro-ph/0402512.}
\nref\fs{W.~Fischler and L.~Susskind, hep-th/9806039; R.~Bousso,
JHEP {\bf 9907} (1999) 004.}
\nref\stq{S. Hellerman, N. Kaloper and L. Susskind, hep-th/0104180.}
\nref\cw{W. Chen and Y. S. Wu, Phys.Rev.D41 (1990) 695.}

The cosmological constant problem is a longstanding problem in
theoretical physics, and has received even more serious
considerations recently, due to the observational evidence for a
non-vanishing value \snc. The direct evidence for the existence of
the dark energy is further supported by other cosmological
observations, in particular by the WMAP experiment \ben. For the
first time in history, theorists are forced to explain not only
why the cosmological constant is small, but also why it is
comparable to the critical density (in this note we shall use
terms the cosmological constant and the dark energy exchangeably.)

A. Cohen and collaborators suggested sometime ago \ckn, that in
quantum field theory a short distance cut-off is related to a long
distance cut-off due to the limit set by formation of a black
hole, namely, if $\rho_\la$ is the quantum zero-point energy
density caused by a short distance cut-off, the total energy in a
region of size $L$ should not exceed the mass of a black hole of
the same size, thus $L^3\rho_\la\le LM_p^2$. The largest $L$
allowed is the one saturating this inequality, thus
\eqn\uvb{\rho_\la=3c^2M_p^2L^{-2}.} For convenience, we introduced
a numerical constant $3c^2$ in the above relation, and use $M_p$
to denote the reduced Planck mass $M_p^{-2}=8\pi G$. Taking $L$ as
the size of the current universe, for instance the Hubble scale,
the resulting energy density is comparable to the present day dark
energy. Related ideas were discussed in \st.

While the magnitude of the holographic energy of Cohen et al.
matches the experimental data, S. Hsu recently pointed out  that
the equation of state does not \sh. Hsu's argument can be refined
as follows. In the Friedman equation $3M_p^2H^2= \rho$, we put two
terms $\rho_m$ and $\rho_\Lambda$, the latter being given by \uvb,
with $L=H^{-1}$. We find \eqn\mattd{\rho_m=3(1-c^2)M_p^2H^2,} thus
$\rho_m$ behaves as $H^2$, the same as $\rho_\la$. But $\rho_m$
scales with the universe scale factor $a$ as $a^{-3}$, so does
$\rho_\la$, thus the dark energy is pressureless, namely in the
equation of state $p=w\rho$, $w=0$. The accelerating universe
certainly requires $w<-1/3$, and the most recent data indicate
that $w<-0.76$ at the $95\%$ confidence level \nsnc.

To remedy the situation, we are forced to use a different scale
other than the Hubble scale as the infrared cut-off. One
possibility quickly comes to mind, the particle horizon used in
the holographic cosmology of Fischler and Susskind \fs. The
particle horizon is given by \eqn\parth{R_H=a\int_0^t{dt\over
a}=a\int_0^a{da\over Ha^2}.} Replacing $L$ in \uvb\ by $R_H$, we
can solve the Friedmann equation exactly with another energy
component (for instance matter). Unfortunately, this replacement
does not work. To see this, we assume that the dark energy
$\rho_\la$ dominates, thus the Friedmann equation simplifies to
$HR_H=c$, or \eqn\parths{{1\over Ha^2}=c{d\over da}({1\over Ha}).}
We find $H^{-1}=\alpha a^{1+{1\over c}}$ with a constant $\alpha$.
The ``dark energy" assumes the form \eqn\parthd{\rho_\la
=3\alpha^2M_p^2a^{-2(1+{1\over c})}.} So $w=-{1\over 3}+{2\over
3c}>-{1\over3}$.

In the relation $HR_H=c$, $c$ is always positive, and in changing
this integral equation into a differential equation \parths, we
find that the changing rate of $1/(Ha)$ with respect to $a$ is
always positive, namely, the Hubble scale $1/H$ as compared to the
scale factor $a$ always increases. To get an accelerating
universe, we need a shrinking Hubble scale. To achieve this, we
replace the particle horizon by the future event horizon
\eqn\eventh{R_h=a\int_t^\infty{dt\over a}=a\int_a^\infty {da\over
Ha^2}.} This horizon is the boundary of the volume a fixed
observer may eventually observe. One is to formulate a theory
regarding a fixed observer within this horizon.

Again, we assume that the dark energy dominates matter, solving
equation \eqn\fhd{\int_a^\infty{da\over Ha^2}={c\over Ha},} we
have
\eqn\goodn{\rho_\la=3c^2M_p^2R_h^2=3\alpha^2M_p^2a^{-2(1-{1\over
c})},} or \eqn\ggdw{w=-{1\over 3}-{2\over 3c}.} Alas, we do obtain
a component of energy behaving as dark energy. If we take $c=1$,
its behavior is similar to the cosmological constant. If $c<1$,
$w<-1$, a value achieved in the past only in the phantom model. A
smaller $c$ although makes the dark energy smaller for a fixed
event horizon size, it also forces $R_h$ to be smaller by the
Friedmann equation $HR_h=c$, thus the changing rate of $1/(Ha)$
larger. This is the reason why a smaller $c$ makes the universe
accelerate faster.

Theoretically, we are more interested in the case $c=1$. We can
actually give an argument in favor of $c=1$. Suppose the universe
be spatially flat (as the observation suggests), the total energy
within a sphere of radius $R_h$ is ${4\pi\over 3}R_h^3\rho_\la$.
On the other hand, the mass of a black hole of size $R_h$ is
$R_h/(2G)$. Equating these two quantities, we find
\eqn\newuvb{\rho_\la={3\over 8\pi G}R_h^{-2}=3M_p^2R_h^{-2},} it
follows that $c=1$.

Before we consider a more realistic cosmology, let us pause to
discuss causality. Since the event horizon $R_h$, as defined in \eventh\
depends on the future evolution of $a(t)$, it appears that our holographic
dark energy grossly violates causality. Event horizon in the context
of cosmology as well as in that of a black hole is always defined
globally, as the casual structure of space-time is a global thing.
The co-moving time is the intrinsic time of a co-moving observer,
and in a time-dependent background it is not the best time to use
in order to understand causality. Indeed, in the conformal time,
the event horizon is no-longer as acausal as in the co-moving time, as
we shall see shortly. The metric $ds^2=-dt^2 +a^2(t)dx^2$ is rewritten in
the conformal time
\eqn\conft{\eta=\int_\infty^t {dt'\over a(t')},}
as
\eqn\confm{ds^2=a^2(\eta)(-d\eta^2+dr^2+r^2d\Omega_2^2).}
Now, the range of the conformal time has a finite upper limit $0$,
for instance $\eta\in (-\infty, 0)$. Due to this finite upper
limit, a light-ray starts from the origin at the time $\eta$ can
not reach arbitrarily far, thus there is a horizon at $r=-\eta$.
(For a more detailed discussion on the global causal structure
of such a universe, see \stq.)
A local quantum field theory for the observer sitting at the origin
is to be defined within this finite
box. We now imagine that a fundamental theory in this finite
box will results in a zero-point energy which is just holographic
dark energy. Now, the formula $R_h=a(\eta)|\eta|$ no longer
appears acaual. Now, the puzzle transforms into the question
how a fundamental theory can be formulated within a finite box,
this is supposed to be a consequence of cosmological complementarity.

Still, it appears rather puzzling why holographic energy is given
by the time-dependent horizon size, as its definition is global.
We may pose a similar puzzle concerning the Gibbons-Hawking entropy.
If the universe evolves adiabatically, then the potential total
entropy of our universe at time $\eta$ is given by $S(\eta)=\pi R_h^2/l_p^2$,
it superficially violates causality as much as holographic dark
energy does. If one eventually can understand the origin of this
entropy, hopefully we may eventually understand the origin of
holographic dark energy (for a discussion on the connection between
entropy and dark energy, see the second reference of \st.)

Although we argued that $c=1$ is preferred, in what follows we leave
$c$ as an arbitrary parameter. With an additional
energy component, the Friedmann equation can always be solved
exactly. For instance, with matter present, the Friedmann equation reads
\eqn\feq{3M_p^2H^2=\rho_0a^{-3}+3c^2M_p^2R_h^{-2},}
where $\rho_0$ is the value of $\rho_m$ at the present time when $a=1$.
This equation can be rewritten as
\eqn\refeq{\int_a^\infty {da\over Ha^2}=c(H^2a^2-\Omega_m^0H_0^2a^{-1})^{-1/2}.}
We may try to convert the above integral equation to a differential
equation for the unknown function $H$.

However, it proves more convenient to use $\Omega_\la$ as the unknown
function. We have $\Omega_\la =\rho_\la/\rho_c$, where $\rho_c=3M_p^2H^2$.
By definition, $R_h^2=3c^2M_p^2/\rho_\la=c^2/(\Omega_\la H^2)$, or
\eqn\fhin{\int_a^\infty {da\over Ha^2}=\int_x^\infty {dx\over Ha}
={c\over\sqrt{\Omega_\la}Ha},} where $x=\ln a$. Next, we wish to
express $Ha$ in terms of $\Omega_\la$. To this end, we introduce
the matter component $\rho_m=\rho^0_ma^{-3}$. We set $a(t_0)=1$,
and $\rho^0_m$ is the present matter energy density. Now, the
Friedmann equation is simply $1-\Omega_\la=\Omega_m
=\Omega_m^0H_0^2H^{-2}a^{-3}$. This implies \eqn\exph{{1\over
Ha}=\sqrt{a(1-\Omega_\la)}{1\over H_0\sqrt{\Omega_m^0}}.}
Substituting this relation (as implied by the Friedmann equation)
into \fhin \eqn\comfe{\int_x^\infty
\sqrt{a}\sqrt{1-\Omega_\la}dx=c\sqrt{a} \sqrt{{1\over
\Omega_\la}-1}.}
Taking derivative with respect to $x$ in the both
sides of the above relation, and noting that the derivative of
$\sqrt{a}$ is proportional to $\sqrt{a}$, we obtain
\eqn\diffo{{\Omega_\la'\over\Omega_\la^2}=(1-\Omega_\la)({1\over\Omega_\la}
+{2\over c\sqrt{\Omega_\la}}),} where the prime denotes the
derivative with respect to $x$. This equation can be solved
exactly. Before solving the equation, we note that $\Omega'_\la$
is always positive, namely the fraction of the dark energy
increases in time, the correct behavior as we expect. Also, the
expansion of the universe will never have a turning point so that
the universe will not re-collapse, since $\Omega'_\la$ never
vanishes before $\Omega_\la$ reaches its maximal value $1$.

Let $y=1/\sqrt{\Omega_\la}$, the differential equation \diffo\ is
cast into the form \eqn\diffoe{y^2y'=(1-y^2)({1\over c}+\half y).}
This equation can be solved exactly for arbitrary $c$,
we write down the solution for $c=1$ only for illustration
purpose:
\eqn\solu{\ln\Omega_\la -{1\over
3}\ln(1-\sqrt{\Omega_\la}) +\ln(1+\sqrt{\Omega_\la})-{8\over
3}\ln(1+2\sqrt{\Omega_\la})=\ln a +x_0.} If we set $a_0=1$ at the
present time, $x_0$ is equal to the L.H.S. of \solu\ with
$\Omega_\la$ replaced by $\Omega_\la^0$

As time draws by, $\Omega_\la$ increases to $1$, the most
important term on the L.H.S. of \solu\ is the second term, we
find, for large $a$
\eqn\asymo{\sqrt{\Omega_\la}=1-3^{-8}2^3e^{-3x_0}a^{-3}.} Since
the universe is dominated by the dark energy for large $a$, we
have
\eqn\frho{\rho_\la\simeq\rho_c=\rho_m/(1-\Omega_\la)=\rho_m^0a^{-3}/(1-\Omega_\la).}
Thus, using \asymo\ in the above relation
\eqn\fde{\rho_\la=3^82^{-4}e^{3x_0}\rho^0_m.} Namely, the final
cosmological constant is related to $\rho_m^0$ through the above
relation.

For very small $a$, matter dominated, and the most important term
on the L.H.S. of \solu\ is the first term, we find
\eqn\erlo{\Omega_\la=e^{x_0}a,} thus
\eqn\erld{\rho_\la=\Omega_\la\rho_c\simeq
\Omega_\la\rho_m=e^{x_0}\rho_m^0a^{-2}.} So although the dark
energy is larger for smaller $a$, it is still dominated over by
matter, we do not have to worry about the possibility of ruining
the standard big bang theory. A discussion of the dark energy
behaving as $a^{-2}$ in the early universe can be found in \cw.

What we are interested in most is the prediction about the
equation of state at the present time. Usually, in the cosmology
literature such as \nsnc, one measures $w$ as in $\rho_\la\sim
a^{-3(1+w)}$. Expanding \eqn\expo{\ln\rho_\la =\ln
\rho_\la^0+{d\ln\rho_\la \over d\ln a} \ln a +\half
{d^2\ln\rho_\la \over d(\ln a)^2}(\ln a)^2+\dots,} where the
derivatives are taken at the present time $a_0=1$. The index $w$
is then \eqn\indx{w=-1-{1\over 3}\left({d\ln\rho_\la \over d\ln a}
+\half {d^2\ln\rho_\la \over d(\ln a)^2}\ln a\right),} up to the
second order. Since $\rho_\la \sim \Omega_\la H^2 \sim \Omega_\la
{\rho_m\over\Omega_m}\sim\Omega_\la/(1-\Omega_\la) a^{-3}$, the
derivatives are easily computed using \diffo: \eqn\spei{w=-{1\over
3}-{2\over 3c}\sqrt{\Omega_\la^0} +{1\over
6c}\sqrt{\Omega_\la^0}(1-\Omega_\la^0)(1+{2\over c}\sqrt{\Omega_\la^0})z,}
where we used $\ln a=-\ln(1+z)\simeq -z$.

The above formula is valid for arbitrary $c$. Specifying to the case
$c=1$ when the holographic dark energy approaching to a constant in the
far future, and plugging the optional
value $\Omega_\la^0=0.73$ into \spei, \eqn\predi{w=-0.903+0.104z.}
Of course only the first two digits are effective. This result is
in excellent agreement with new data \nsnc. At the one sigma
level, the result of \nsnc\ is $w=-1.02^{+0.13}_{-0.19}$, with a
slightly different value $\Omega_\la^0=0.71$. If our holographic
model for dark energy is viable, it is quite hopeful that this
prediction will be verified by experiments in near future.

The choice $c<1$ will leads to dark energy behaving as phantom, and in this
case, the Gibbons-Hawking entropy will eventually decrease as the
the event horizon will shrink, this violates the second law of
thermodynamics. For $c>1$, the second law of thermodynamics is
not violated, while in a situation without any other component
of energy, the space-time is not de Sitter, thus for symmetry reason
we prefer to choose $c=1$ and the result \predi\ in a sense is
a prediction.

During the radiation dominated epoch, the dark energy also
increases with time compared to the radiation energy, but it is
still small enough not to ruin standard results such as nuclear
genesis. We are also interested in whether our model will greatly
affect the standard slow-roll inflation scenario. In this case,
assume that the universe has only two energy components, the
``dark energy" and the inflaton energy. If the latter is almost
constant, we shall show that it is possible that the dark energy
can be inflated away. Similar to \diffo, in this case we can
derive an equation
\eqn\inder{\Omega_\la'=2\Omega_\la(\Omega_\la-1)(1-\sqrt{\Omega_\la}).}
Thus, $\Omega_\la$ always decreases during inflation. The above
equation can also be solved exactly. Instead of exhibiting the
exact solution, we only show its behavior for small $\Omega_\la$:
\eqn\redsh{\Omega_\la\sim a^{-2},} thus, if the initial value of
$\Omega_\la$ is reasonable, it will be red-shifted away quickly
enough not to affect the standard inflation scenario.

This huge red-shift may be the resolution to the cosmic
coincidence problem, since the coincidence problem becomes a
problem of why the ratio between the dark energy density and the
radiation density is a very tiny number at the onset of the
radiation dominated epoch. A rough estimate shows that the ratio
between $\rho_\la$ and $\rho_r$, the radiation density, is about
$10^{-52}$, if we choose the inflation energy scale be
$10^{14}$Gev. According to $\redsh$, this is to be equal to $\exp
(-2N)$ where $N$ is the number of e-folds, and we find  $N=60$,
the minimal number of e-folds in the inflation scenario. Of
course, we need to assume that all the dark energy is included in
$\rho_\la$ in the end of inflation, namely, the inflaton energy
completely decayed into radiation. Thus, inflation not only solves
the traditional naturalness problems and helps to generate
primordial perturbations, it also solves the cosmic coincidence
problem! We may imagine that in another region of the universe,
the number of e-folds is different, thus a different cosmological
constant results.

This model requires, for a consistent solution to exist, that
any other form of energy must eventually decay. Still, it is possible
that there is an additional component of dark energy such as
quintessence which will indeed decay in the far future. A couple
of papers explored this question after the present paper appeared
on the internet, so we shall not address this question here.

In conclusion, the holographic dark energy scenario is viable if
we set the infrared cut-off by the event horizon. This is not only
a viable model, it also makes a concrete prediction about the
equation of state of the dark energy, thus falsifiable by the
future experiments.

However, unlike expected earlier, we are not able to explain the
cosmic coincidence along the line of \ckn, since the infrared
cut-off is not the current Hubble scale. The eventual cosmological
constant in the far future can be viewed as a boundary condition,
or equivalently, the initial value of $\Omega_\la$ can be viewed
as a initial condition. This initial condition is affected by
physics in very early universe, for instance physics of inflation.
In this regard, it appears that inflation is able to explain the
current value $\rho_\la$ if a proper number of e-folds is assumed,
since the dark energy compared to the inflaton energy thus the
radiation energy in the end of inflation is very small due to
inflation. A detailed analysis will appear elsewhere.

Acknowledgments.

I am grateful to Q. G. Huang, J. X. Lu and X. J. Wang, for useful
discussions. This work was supported by a ``Hundred People
Project'' grant of Academia Sinica and an outstanding young
investigator award of NSF of China. This work was done during a
visit to the Interdisciplinary Center for Theoretical Study at
University of Science and Technology of China.

\listrefs
\end